\documentclass{appolb}
\pdfoutput=1

\usepackage{graphicx}
\newcommand{\herwig}{\textsc{Herwig}}
\newcommand{\pythia}{\textsc{Pythia}}
\newcommand{\sherpa}{\textsc{Sherpa}}
\newcommand{\ptmin}{p_t^{\rm min}}

\newcommand{\Sinc}{\sigma^{\rm inc}}
\newcommand{\diff}[2]{\frac{\mathrm{d}#1}{\mathrm{d}#2}}
\newcommand{\dd}{\mathrm{d}}
\newcommand{\pcr}{PCR}
\newcommand{\scr}{SCR}
\newcommand{\ptlead}{\ensuremath{p_{\perp}^{\rm lead}}}

\begin{document}
\title{
\vspace{-4.0cm}
\begin{flushright}
		{\small {\bf KA-TP-05-2013}}   \\
		{\small {\bf MAN/HEP/2013/03}}  \\
		{\small {\bf MCnet-13-02}}  \\
\end{flushright}
\vspace{0.5cm}
Multiple Partonic Interactions in \herwig++\thanks{Presented by 
Andrzej Si\'odmok at 42. International Symposium on Multiparticle 
Dynamics (ISMD 2012), Kielce, Poland}%
}
\author{Stefan Gieseke, Christian R\"ohr
\address{Institut f\"ur Theoretische Physik, Karlsruhe Institute of Technology
           (KIT), Karlsruhe, Germany}\\
\vspace*{0.5cm}
{
}
Andrzej Si\'odmok
\address{ Consortium for Fundamental Physics, School of Physics and Astronomy, 
\\The University of Manchester, Manchester, U.K. }
}
\maketitle
\begin{abstract}
We review the implementation of a model for multiple partonic interactions in 
\herwig{}++. Moreover, we show how recent studies on the colour structure of 
events in \herwig{}++ led to a significant improvement in the description of 
soft inclusive observables in pp interactions at the LHC.
\end{abstract}
  
\section{Introduction}
Multiple partonic interaction (MPI) models are crucial for a successful description 
of the underlying event (UE) in hard hadronic collisions and minimum-bias (MB) 
data from the Tevatron and the Large Hadron Collider (LHC). The modern Monte Carlo 
event generators \herwig{++}~\cite{Bahr:2008pv}, \pythia~\cite{Sjostrand:2006za,Sjostrand:2007gs} 
and \sherpa~\cite{Gleisberg:2008ta} all include an MPI model in order to simulate 
the underlying event. In this short note we summarize the modelling of multiple parton interactions in the 
\herwig++{} event generator, present its recent developments and show a comparison
of the improved model to LHC data.
\section{MPI model in \herwig{++}}
The modelling of underlying events in \herwig{}++ is based on the fact that at high 
enough energies the hard inclusive cross section for dijet production,
\begin{equation}
  \Sinc(s;\ptmin) = \sum_{i,j} \int_{{\ptmin}^2} {\dd p_t^2}
  f_{i/h_1}(x_1, \mu^2) \otimes \diff{\hat\sigma_{i,j}}{p_t^2} \otimes
  f_{j/h_2}(x_2, \mu^2) \, ,
  \label{eq:sigmainc}
\end{equation}
will eventually exceed the total cross section \cite{Bahr:2008wk}.
This leads to the interpretation that in fact the inclusive cross section counts 
not only single hard events but all hard events that occur in parallel during the 
very same hadron-hadron collision.
\subsection{Eikonal model}
With the eikonal assumption that at fixed impact parameter, $\vec b$, individual 
scatterings are independent\footnote{Of course some basic correlations 
are incorporated into the model later on, for example by requirement of momentum 
conservation.} and that the distribution of partons in hadrons factorizes with respect 
to the $\vec b$ and $x$ dependence, we can express the average number of hard interactions 
in a hadron-hadron collision as:
\begin{equation}
  \bar n(\vec b, s) = A(\vec b; \mu^2) \Sinc (s; \ptmin)\ ,
\end{equation}
where the function $A(\vec b; \mu^2)$ describes the spatial overlap of
the two colliding hadrons (protons) as a function of the impact
parameter $\vec b$. In both Fortran \herwig{}~\cite{Corcella:2000bw} with a Jimmy plug-in~\cite{Butterworth:1996zw} 
and \herwig{}++, the overlap function is modelled according to the electromagnetic form 
factor~\cite{{Borozan:2002fk}} and the parameter $\mu$ interpreted as the inverse 
radius of the proton. Since the spatial parton distribution (the colour distribution) is assumed 
to be similar to the distribution of electric charge, but not necessarily identical, $\mu$ 
is treated as a free parameter. This model allows for the simulation of multiple interactions 
with perturbative scatters which have $p_t > \ptmin$.

\begin{figure*}[htb]
  \includegraphics[width=0.45\textwidth]{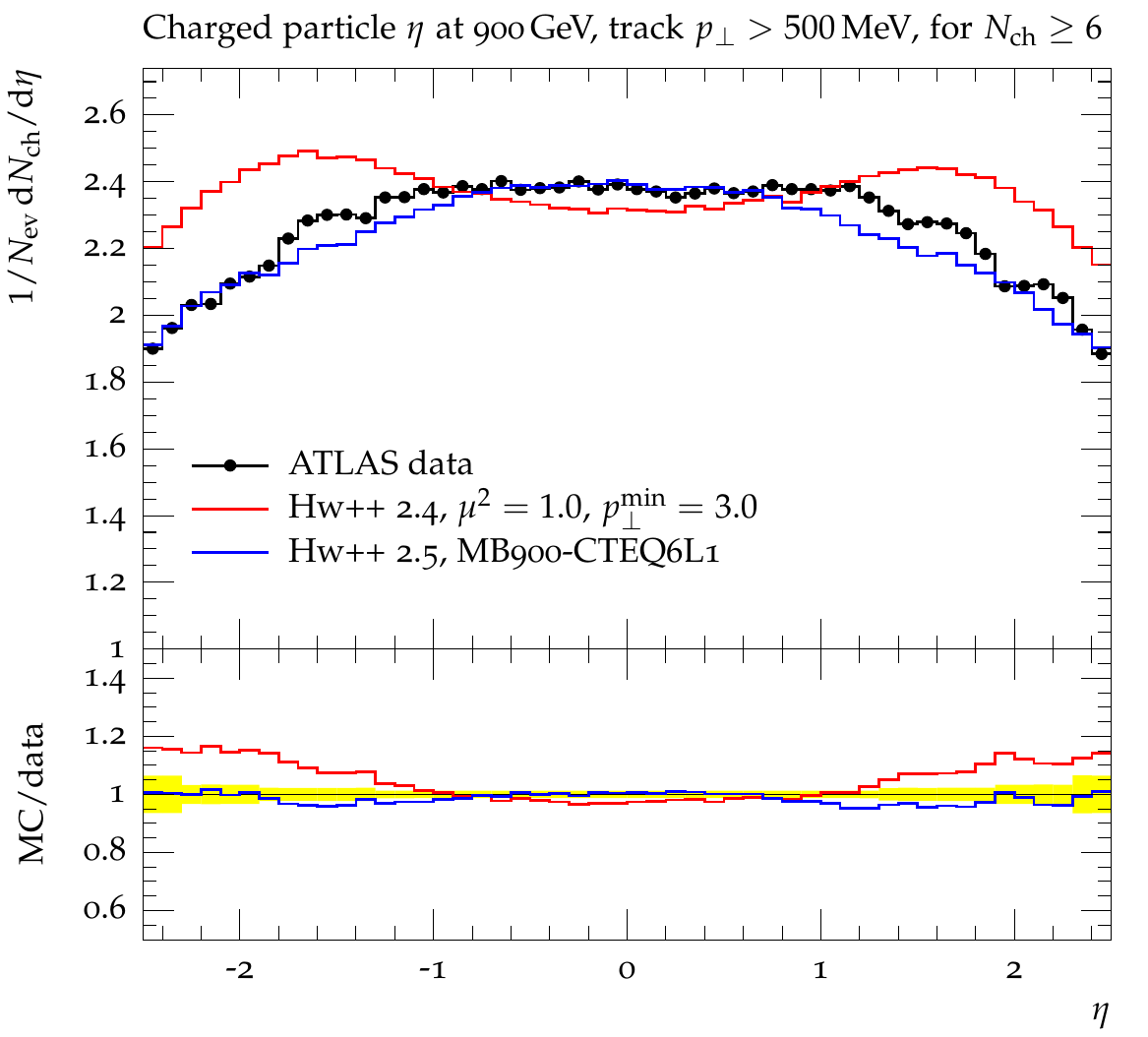}
   \includegraphics[width=0.45\textwidth]{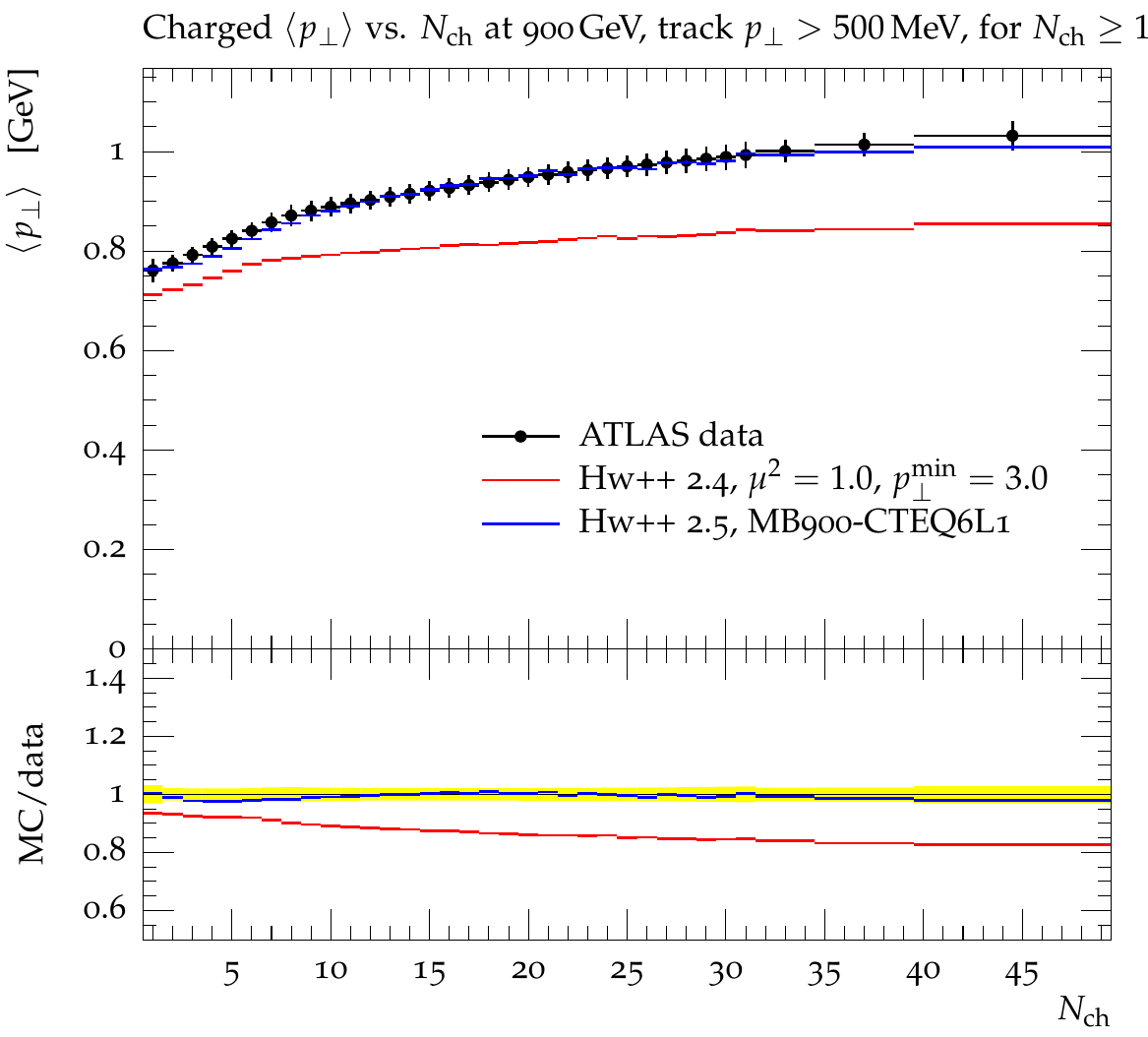}\\
   \caption{Comparison of \herwig{} 2.4.2 without CR and \herwig{} 2.5 with \pcr{} to ATLAS minimum-bias 
   distributions at $\sqrt{s}=0.9${TeV}. 
}
  \label{fig:ATLAS_900_Nch6}
\end{figure*}

The extension to soft scatterings (with $p_t < \ptmin$) is kept as simple as possible. 
The additional soft contribution to the inclusive cross section is also eikonalized, such 
that we can also calculate an average number of soft scatters from the resulting 
$\Sinc_{\rm soft}$ and an overlap function $A_{\rm soft}(\vec b)$ for the soft scattering 
centres.  The functional form $A_{\rm soft}(\vec b)$ is assumed to be the same as for
the hard scatters, but we allow for a different inverse radius, $\mu_{\rm soft}$.  
The consistency of this model with unitarity is given by fixing the two additional parameters 
$\Sinc_{\rm soft}$ and $\mu_{\rm soft}^2$ from two constraints.  First, we can calculate the 
total cross section from the eikonal model and fix it to be consistent with the Donnachie-Landshoff 
parametrization~\cite{Donnachie:1992ny}.  In addition, using the optical theorem, we can
calculate the $t$--slope parameter from the eikonal model and fix it to a reasonable 
parametrization. This model and its extension to soft scatterings were implemented in 
\herwig++{}~\cite{Bahr:2008wk,Bahr:2009ek} and proved to be capable of describing the whole spectrum 
of UE data from the Tevatron~\cite{Affolder:2001xt,Acosta:2004wqa} (including its minimum bias part).
However, the MPI model is too simple to describe the first MB data from ATLAS~\cite{Aad:2010rd,Aad:2010ac}
(see red line in Fig.~\ref{fig:ATLAS_900_Nch6}).
A tuning of the parameters of the model~\cite{Bartalini:2011jp} did not improve the description. 
This observation led to new developments of the MPI model to include non-perturbative colour 
reconnections (CR)~\cite{Gieseke:2011xy,Gieseke:2012ft}. 
\subsection{Colour reconnection}
\begin{figure*}[htb]
  \resizebox{1.0\hsize}{!}{
  \begin{minipage}[t]{0.5\textwidth}
    \includegraphics[width=0.7\textwidth]{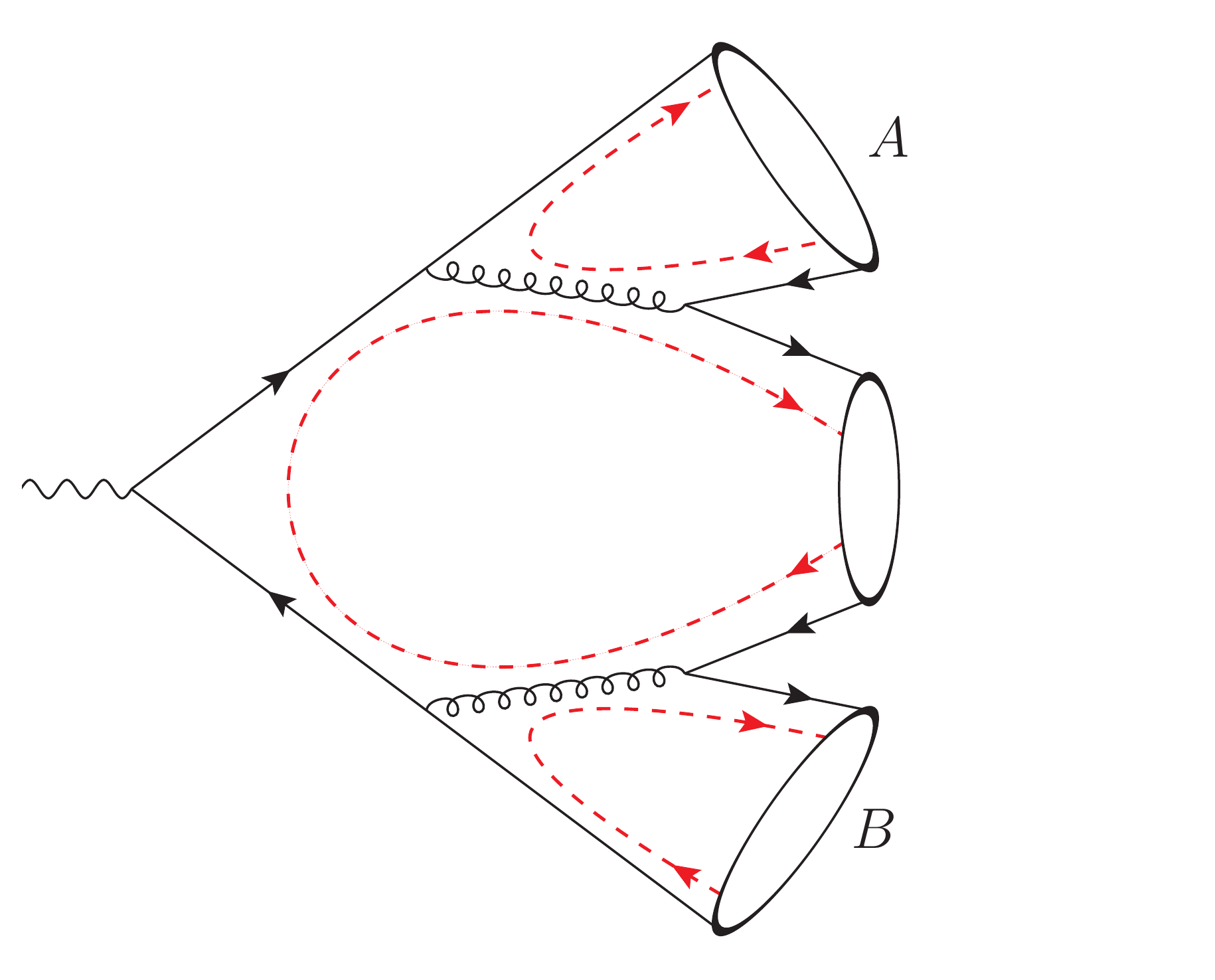}
  \end{minipage}
  \begin{minipage}[t]{0.5\textwidth}
    \includegraphics[width=0.7\textwidth]{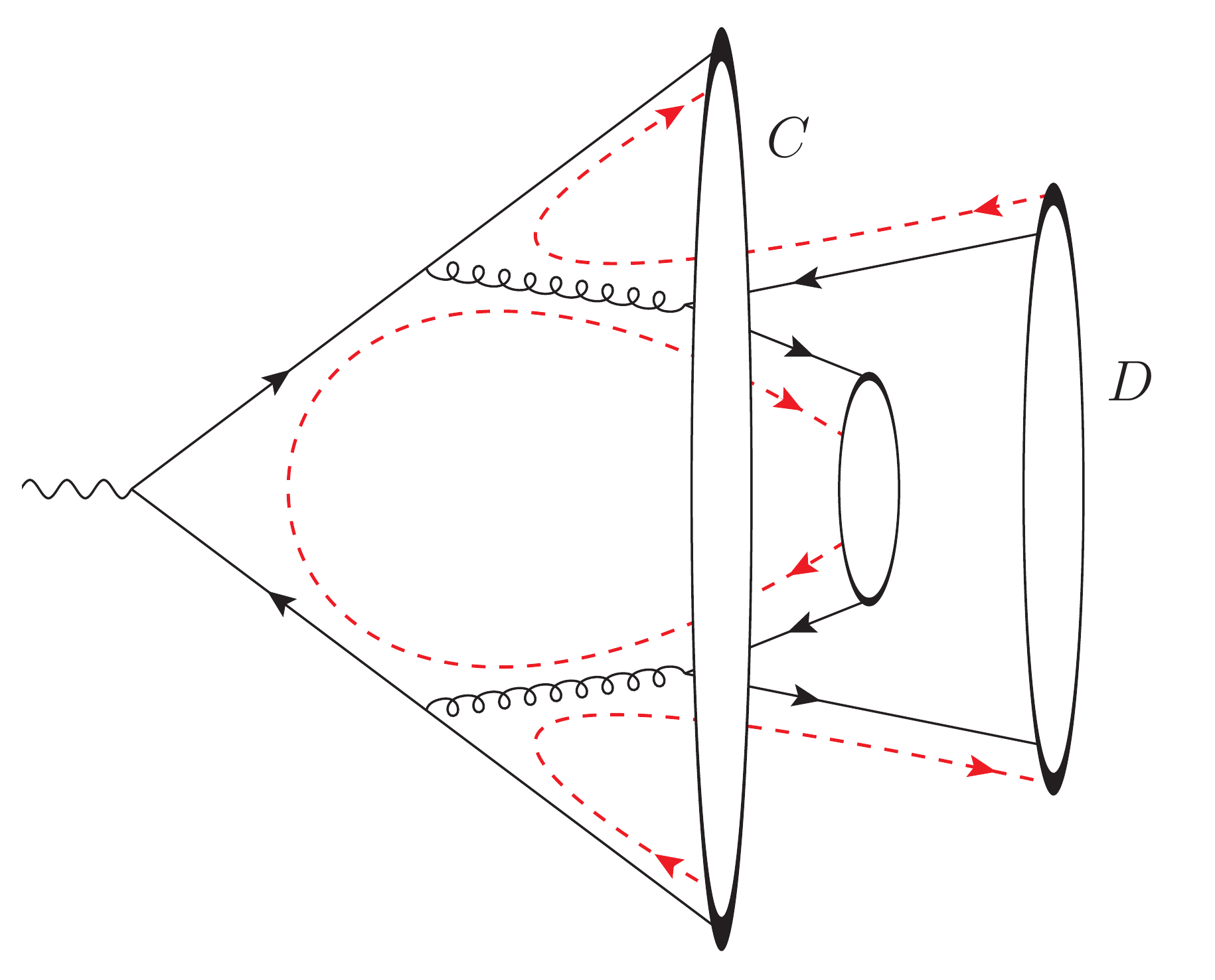}
  \end{minipage}
  }
  \caption{Formation of clusters, which we represent by ovals here. Colour lines
  are dashed.
  The {right diagram} shows a possible colour-reconnected state: the
  partons of the clusters $A$ and $B$ are arranged in new clusters, $C$ and
  $D$.}
  \label{fig:crsketches}
\end{figure*}

The colour structure of multiple interactions can cause non-trivial 
chang\-es to the colour topology of the colliding system as a whole, with potentially 
major consequences for the particle multiplicity in the final state.
The colour connections between partons define colour singlet objects known as 
clusters (see Fig.~\ref{fig:crsketches}). 
The cluster hadronization model~\cite{Webber1984492}, which is used in \herwig{++}, generates clusters
which are used as the starting point for the generation of hadrons via cluster decays. The idea of CR is based 
on colour preconfinement~\cite{Amati:1979fg}, which implies that parton jets emerging from different 
partonic interactions are colour-connected (clustered) if they are located closely in phase space. As the  
MPI model does not take that into account, those colour connections have to be adapted afterwards 
by means of a CR procedure pictured in Fig.~\ref{fig:crsketches}. In our CR model we define the distance between 
two partons to be small when their invariant mass (cluster mass) is small. Therefore, the aim of the CR model is to
reduce the colour length $\lambda \equiv \sum_{i=1}^{N_{cl}}m_i^2$, where $N_{cl}$ is the number of clusters 
in an event and $m_i$ is the invariant mass of cluster~$i$.
Based on this physical motivation, there are two colour reconnection models implemented in \herwig{}++. 
The first CR model, so-called plain CR (\pcr), has been included in \herwig{++} as of version 2.5~\cite{Gieseke:2011na}. 
This model iterates over all cluster pairs in a random order. Whenever a swap of colours is preferable,
i.e.\ when the new cluster masses are smaller, this is done with a given probability (the only parameter of the model).
The \pcr{} model has been shown to give the desired results (see blue line in Fig.~\ref{fig:ATLAS_900_Nch6}). 
However, as the clusters are presented to the model only in a given sequence, it is hard to assess which clusters
are affected and to what extent the sequence is physically relevant. Therefore, we implemented another model, 
the statistical CR model (\scr)~\cite{Gieseke:2012ft}, which adopts the Metropolis algorithm to reduce $\lambda$.
This allowed us to systematically study the consequences of the introduction of CR, the results of which are presented
in details in~\cite{Gieseke:2012ft,Gieseke:2012pt}. 
\section{Tuning and Results}
\begin{figure*}[htb]
  \includegraphics[width=0.32\textwidth]{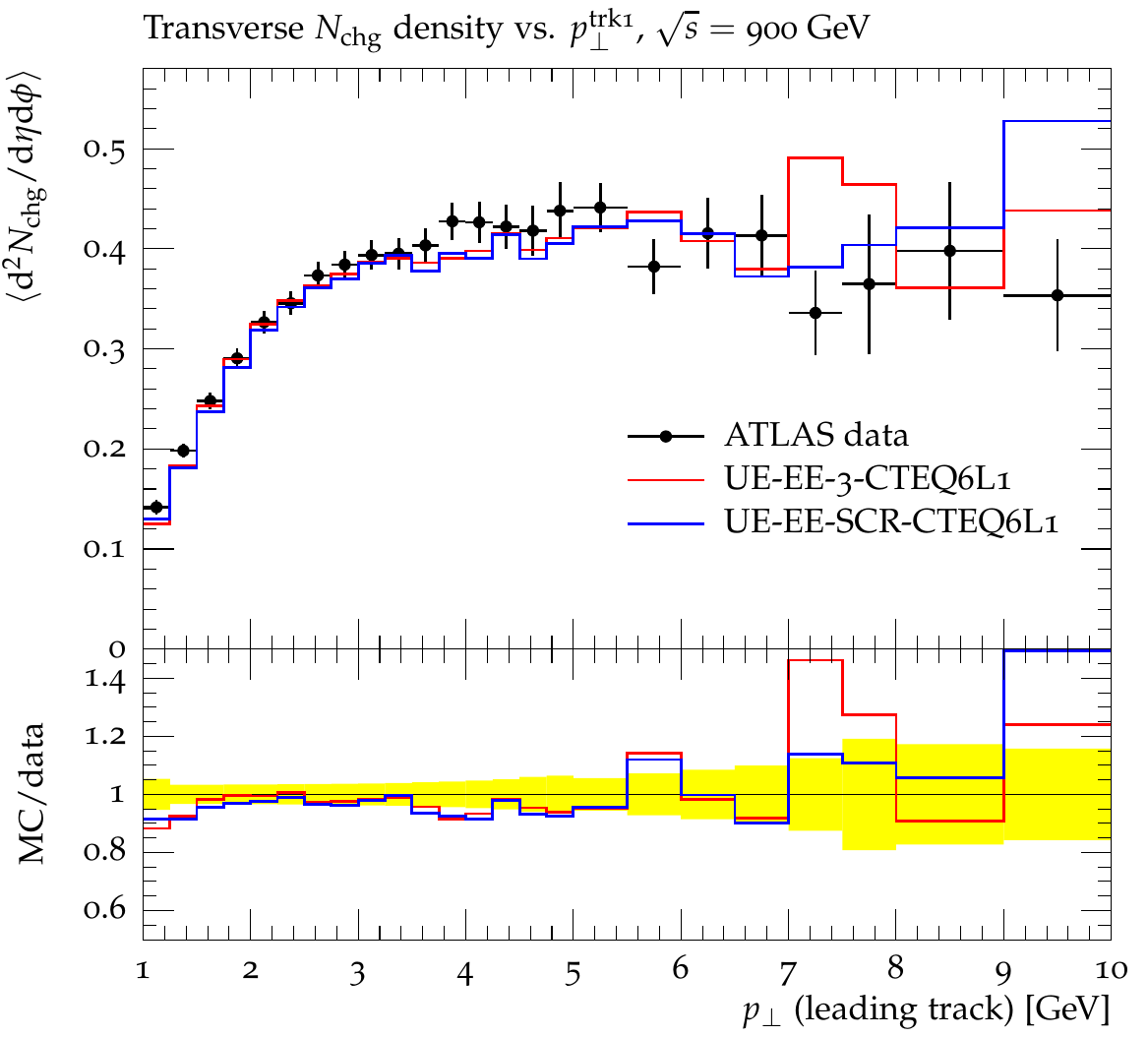}
  \includegraphics[width=0.32\textwidth]{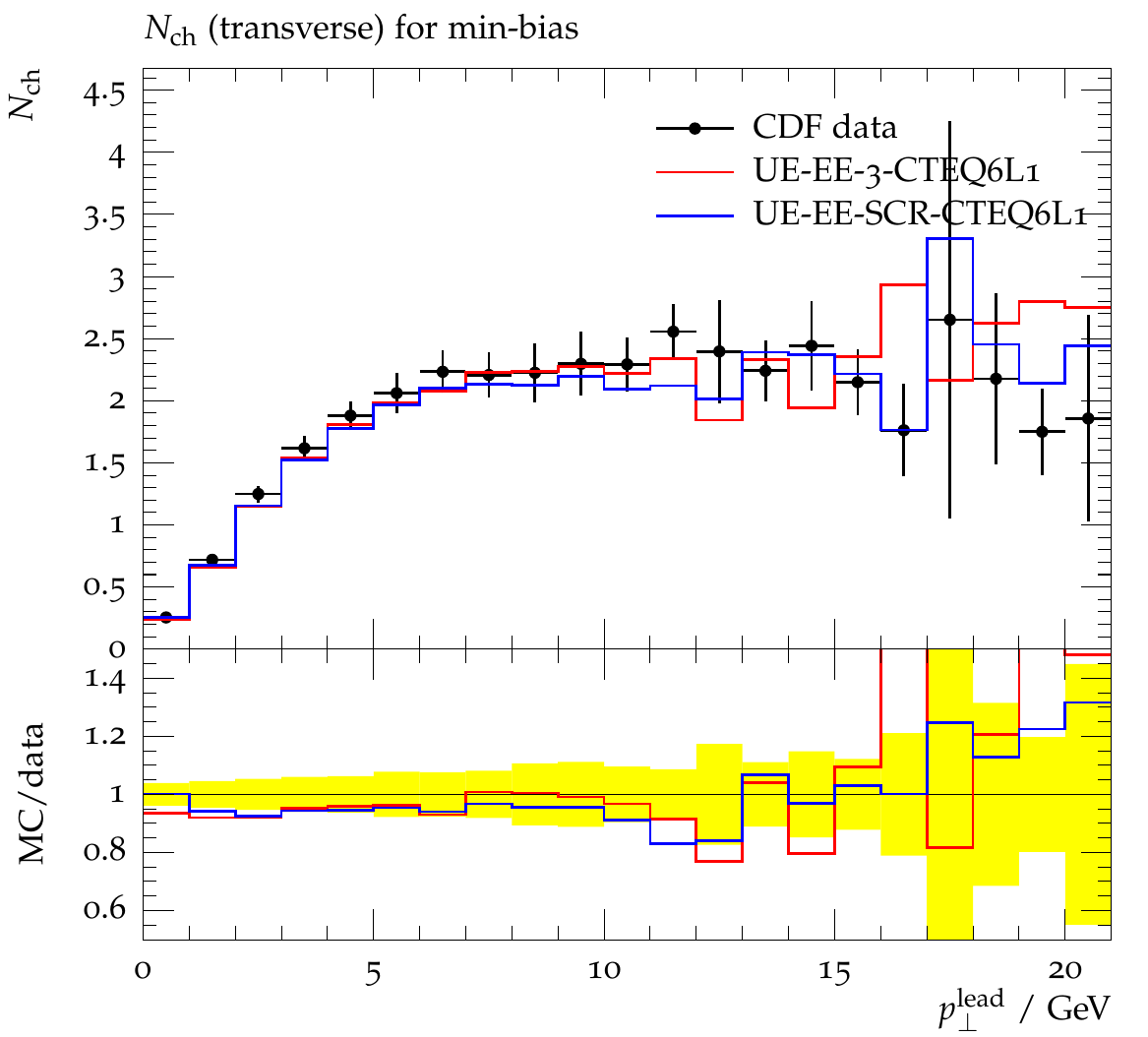}
  \includegraphics[width=0.32\textwidth]{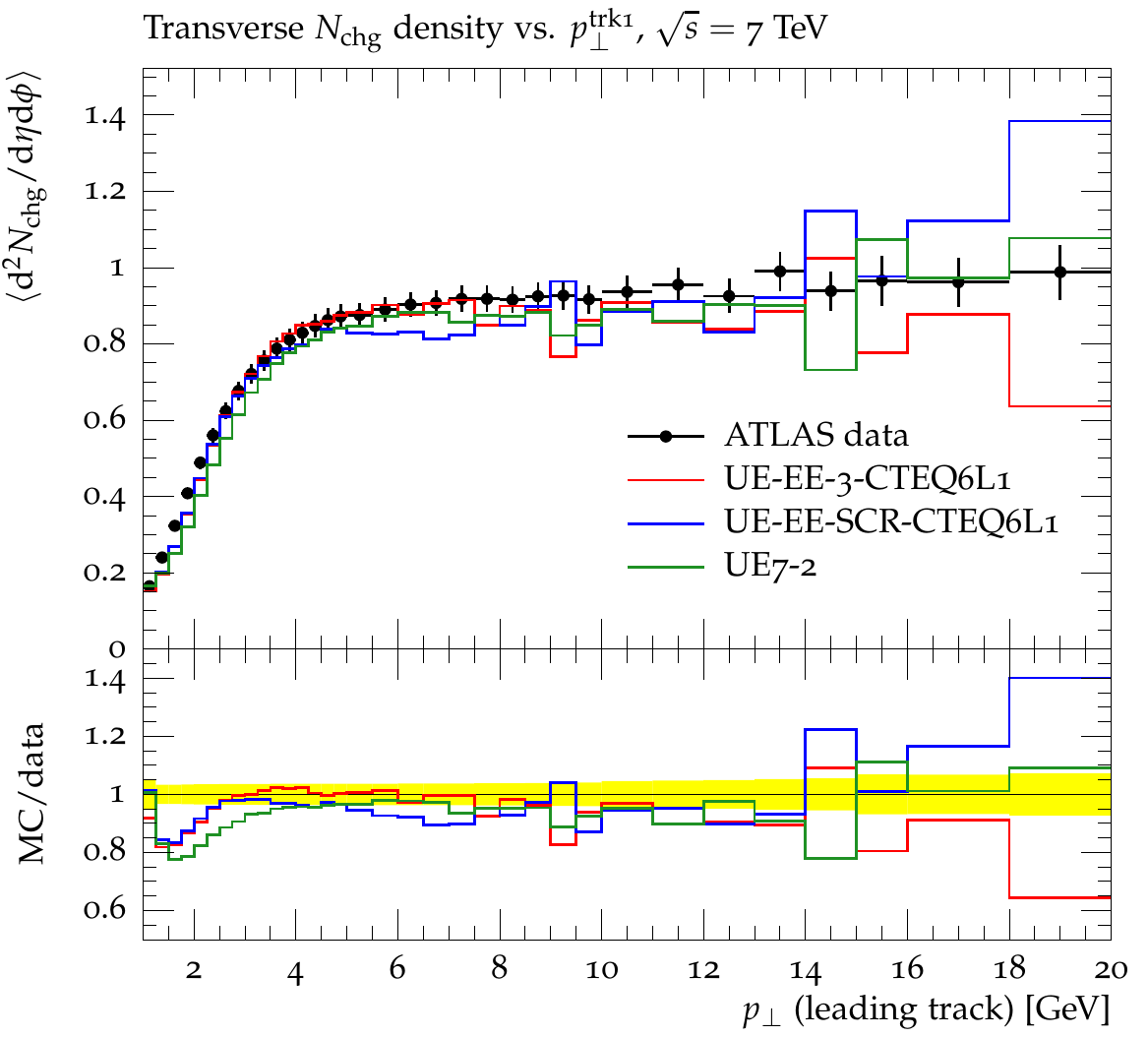}
  \caption{
ATLAS data at ${900}${GeV} (1st column), CDF data at
  ${1800}${GeV} (2nd column) and ATLAS data at ${7}${TeV} (3rd column),
  showing the multiplicity density  of the charged particles in the
 ``transverse'' area as a function of $\ptlead$. 
}
  \label{fig:UE-comparison-transverse}
\end{figure*}

\begin{figure*}[htb]
  \resizebox{1.0\hsize}{!}{
  \begin{minipage}[t]{0.5\textwidth}
    \includegraphics[width=0.85\textwidth]{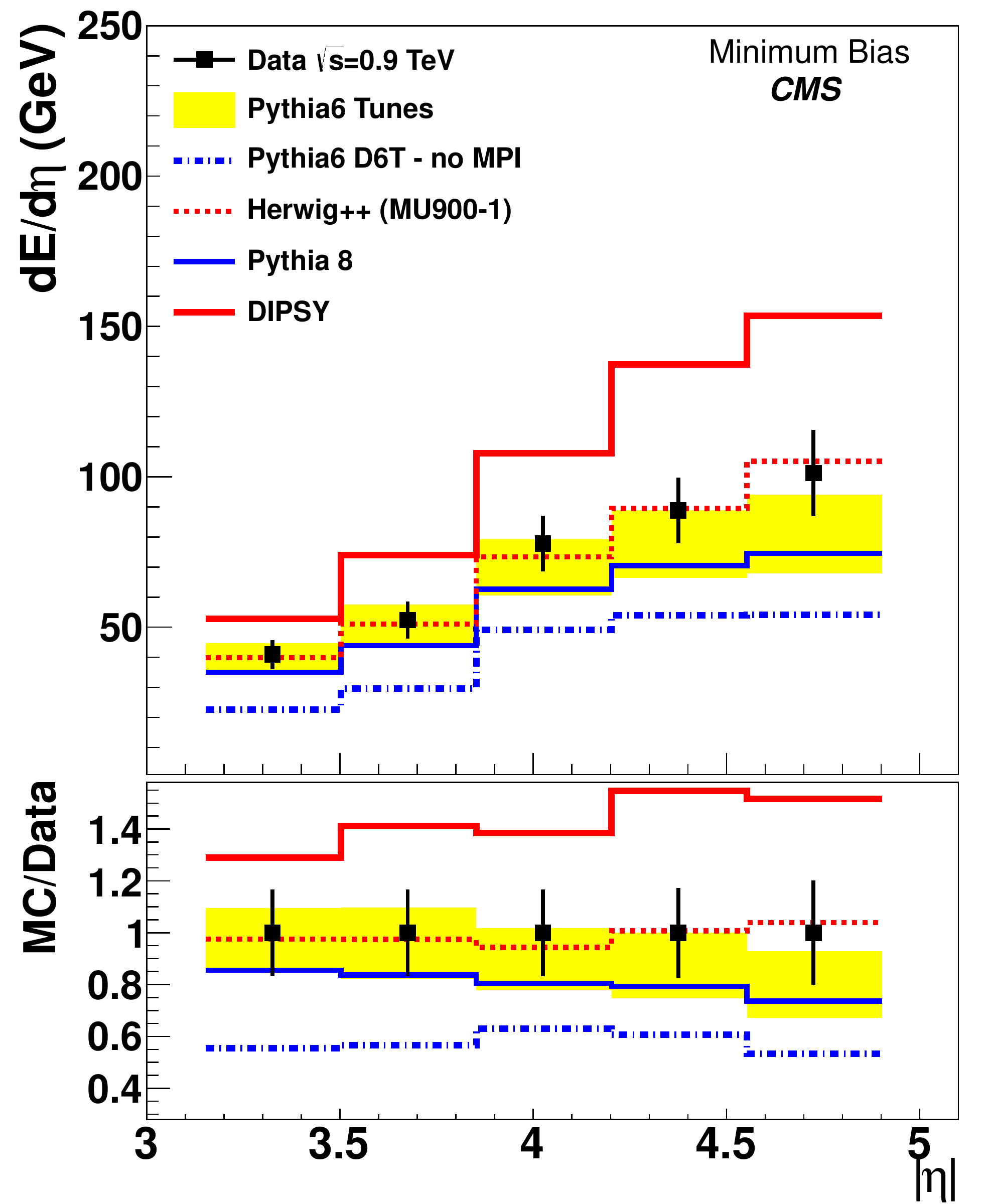}
  \end{minipage}
  \begin{minipage}[t]{0.5\textwidth}
    \includegraphics[width=0.85\textwidth]{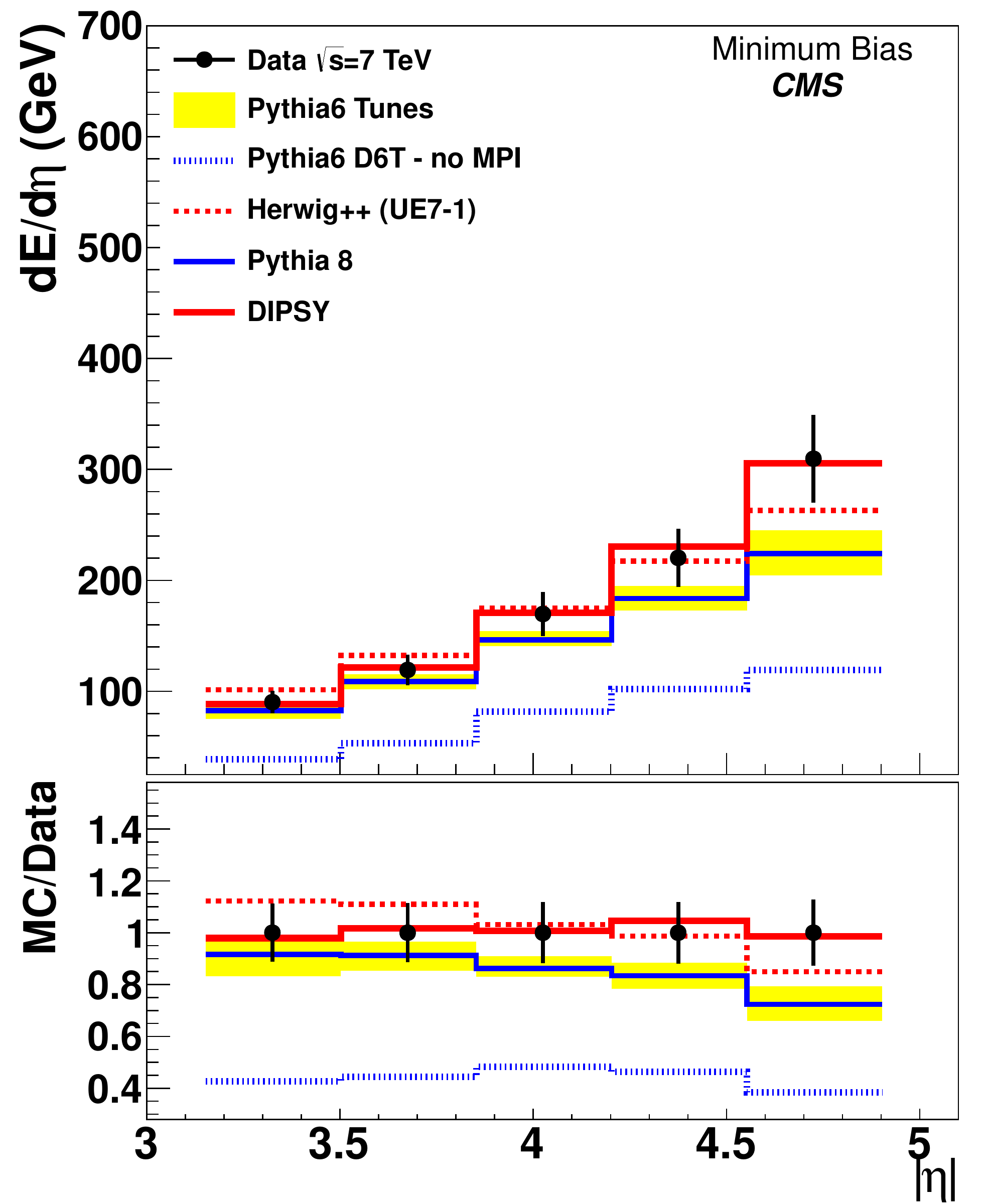}
  \end{minipage}
  }
  \caption{Energy flow  as a function of $\eta$ for MB events at $\sqrt{s}  = 0.9$ and 7~TeV.}
  \label{fig:cms}
\end{figure*}

Initially we were primarily aiming at an improved description of MB data, therefore we started by tuning 
the PCR model to ATLAS MB data\footnote{Since currently there is no
model for soft diffractive physics in \herwig{++}, we use the diffraction-reduced ATLAS MB measurement with
an additional cut on the number of charged particles, Nch $\ge 6$.} (results are shown in Fig.~\ref{fig:ATLAS_900_Nch6}).
The next important  question was whether the new model is able to describe the UE data sets at different 
collider energies. A dedicated tuning procedure~\cite{Gieseke:2008ep} resulted in energy-extrapolated tunes UE-EE, 
for both \pcr{} and \scr{}, in which all parameters are fixed except for $\ptmin$, which varies with energy. 
These tunes allow us to describe data at different energies~\cite{Affolder:2001xt,Aad:2010fh} 
(see Fig.~\ref{fig:UE-comparison-transverse}) with a simple parametrization of the $\ptmin$ energy dependence. 
This was an important step towards the understanding of the energy dependence of the model. 
Finally, we show the results of the model for energy flow as a function of $\eta$ 
for minimum-bias events at $\sqrt{s} = 0.9$ and 7~TeV (Fig.~\ref{fig:cms}) presented by the CMS collaboration 
during this workshop and also published in~\cite{Chatrchyan:2011wm}. 
This observable was not available during the preparation of the tunes and 
these very good results can therefore be treated as a prediction of the model.
The new model is implemented and available with the recent tunes in \herwig{++} version 2.6 \cite{Arnold:2012fq}.
\section{Summary}
We have summarized the multiple partonic interaction models in \herwig{++}  and expanded
on the motivation and modelling of colour reconnection. Furthermore, we have shown that
(sufficiently diffraction-suppressed) MB data from the LHC and underlying-event
observables are well described by the present model which makes it useful for the LHC
collaborations.

\section{Acknowledgements}
We thank the organizers for the very pleasant and fruitful workshop. This work has been 
supported by the Helmholtz Alliance ''Physics at the Terascale`` and by the 
Lancaster-Manchester-Sheffield Consortium for Fundamental Physics under STFC grant ST/J000418/1. 
We wish to acknowledge Dermot Moran for his critical reading of 
this proceedings.
\begin{footnotesize}

\begin{thebibliography}{99}
\bibitem{Bahr:2008pv}
  M.~B\"ahr {\it et al.},
  Eur.\ Phys.\ J.\  C {\bf 58} (2008) 639


\bibitem{Sjostrand:2006za}
  T.~Sj\"ostrand, S.~Mrenna and P.~Z.~Skands,
  JHEP {\bf 0605} (2006) 026

\bibitem{Sjostrand:2007gs}
  T.~Sj\"ostrand, S.~Mrenna, P.~Z.~Skands,
  Comput.\ Phys.\ Commun.\  {\bf 178} (2008)

\bibitem{Gleisberg:2008ta}
  T.~Gleisberg, S.~Hoeche, and F.~Krauss {\it et al.}
  JHEP {\bf 0902} (2009) 007
\bibitem{Bahr:2008wk}
  M.~B\"ahr, J.~M.~Butterworth and M.~H.~Seymour,
  JHEP {\bf 0901} (2009) 065


\bibitem{Corcella:2000bw}
  G.~Corcella, I.~G.~Knowles, G.~Marchesini, S.~Moretti, K.~Odagiri, P.~Richardson, M.~H.~Seymour and B.~R.~Webber,
  JHEP {\bf 0101} (2001) 010
\bibitem{Butterworth:1996zw}
  J.~M.~Butterworth, J.~R.~Forshaw and M.~H.~Seymour,
  Z.\ Phys.\ C {\bf 72} (1996)

\bibitem{Borozan:2002fk}
  I.~Borozan and M.~H.~Seymour,
  JHEP {\bf 0209} (2002) 015

\bibitem{Donnachie:1992ny}
  A.~Donnachie and P.~V.~Landshoff,
  Phys.\ Lett.\ B {\bf 296} (1992) 227
\bibitem{Bahr:2009ek}
  M.~B\"ahr, J.~M.~Butterworth, S.~Gieseke and M.~H.~Seymour,
   arXiv:0905.4671.

\bibitem{Affolder:2001xt}
  T.~Affolder {\it et al.}  [CDF Collaboration],
  Phys.\ Rev.\ D {\bf 65} (2002) 092002.

\bibitem{Acosta:2004wqa}
  D.~Acosta {\it et al.}  [CDF Collaboration],
  Phys.\ Rev.\ D {\bf 70} (2004) 072002
\bibitem{Aad:2010rd}
  G.~Aad {\it et al.}  [ATLAS Collaboration],
  Phys.\ Lett.\ B {\bf 688} (2010) 21
\bibitem{Aad:2010ac}
  G.~Aad {\it et al.}  [ATLAS Collaboration],
  New J.\ Phys.\  {\bf 13} (2011) 053033

\bibitem{Bartalini:2011jp}
  P.~Bartalini, E.~L.~Berger, B.~Blok and G.~Calucci, {\it et al.},
   arXiv:1111.0469.

\bibitem{Gieseke:2011xy}
  S.~Gieseke, C.~A.~R\"ohr and A.~Si\'odmok,
  arXiv:1110.2675 [hep-ph].

\bibitem{Gieseke:2012ft}
  S.~Gieseke, C.~R\"ohr and A.~Si\'odmok,
  Eur.\ Phys.\ J.\ C {\bf 72} (2012) 2225

\bibitem{Webber1984492}
 B.R.~Webber,
 Nuclear Phys.\ B {\bf 238} (1984) 492

\bibitem{Amati:1979fg}
    D.~Amati, and G.~Veneziano
     Phys.\ Lett.\ {\bf B83} (1979) 87

\bibitem{Gieseke:2011na}
  S.~Gieseke, D.~Grellscheid, and  K.~Hamilton {\it et al.},
  arXiv:1102.1672 [hep-ph].
\bibitem{Gieseke:2012pt}
  S.~Gieseke, C.~R\"ohr and A.~Si\'odmok,
  arXiv:1206.2205 [hep-ph].
\bibitem{Gieseke:2008ep}
  S.~Gieseke, C.~A.~R\"ohr and A.~Si\'odmok,
  DESY-PROC-2012-03.

\bibitem{Aad:2010fh}
  G.~Aad {\it et al.}  [Atlas Collaboration],
  Phys.\ Rev.\ D {\bf 83} (2011) 112001

\bibitem{Arnold:2012fq}
  K.~Arnold, L.~d'Errico, and S.~Gieseke {\it et al.},
  arXiv:1205.4902 [hep-ph].
\bibitem{Chatrchyan:2011wm}
  S.~Chatrchyan {\it et al.}  [CMS Collaboration],
  JHEP {\bf 1111} (2011) 148

\end{thebibliography}

\end{footnotesize}
\end{document}